\documentclass[pre,aps,preprint,showpacs,preprintnumbers,amsmath,amssymb,secnumroman,eqsecnum,eqappnum]{revtex4}

\usepackage{graphicx}% Include figure files
\usepackage{dcolumn}% Align table columns on decimal point
\usepackage{bm}% bold math

\newcommand{\beq}{\begin{equation}}
\newcommand{\eeq}{\end{equation}}
\newcommand{\beqa}{\begin{eqnarray}}
\newcommand{\eeqa}{\end{eqnarray}}
\newcommand{\vc}[1]{\mbox{\boldmath $#1$}}

\newcommand{\vol}[1]{{\bf #1}}

\begin{document}

%\preprint{APS/123-QED}

\title{Second harmonic generation and vortex shedding by a dipole-quadrupole and a quadrupole-octupole swimmer in a viscous incompressible fluid}% Force line breaks with \\

\author{B. U. Felderhof}

 \email{ufelder@physik.rwth-aachen.de}
\affiliation{Institut f\"ur Theorie der Statistischen Physik\\ RWTH Aachen University\\
Templergraben 55\\52056 Aachen\\ Germany\\
}%

\author{R. B. Jones}

 \email{r.b.jones@qmul.ac.uk}
\affiliation{Queen Mary University of London, The School of
Physics and Astronomy, Mile End Road, London E1 4NS, UK\\}%

\date{\today}% It is always \today, today,

\begin{abstract}
Vortex shedding by a swimming sphere in a viscous incompressible fluid is studied for surface modulation characterized by a superposition of dipolar and quadrupolar, as well as for quadrupolar and octupolar displacements, varying harmonically in time. The time-dependent swimming velocity and the flow velocity are calculated to second order in the amplitude of surface modulation for both models. The models are also useful for the discussion of bird flight.
\end{abstract}

\maketitle
\section{\label{I}Introduction}

Up till now a fully satisfactory theory of the swimming of fish and the flying of birds is missing. Existing theory is based on the Euler equations of inviscid flow with the effect of viscosity relegated to a boundary layer \cite{1}-\cite{4}. The flow is predominantly irrotational, apart from the boundary layer and a wake of vorticity. It is important to have a simple model of swimming which can be studied in detail. We assume that the Navier-Stokes equations for a viscous incompressible fluid can be used as basic equations describing the dynamics of the fluid. As a model we consider a spherical body which can change its shape periodically in time. The fluid and the body interact via a no-slip boundary condition at the undulating surface of the body. The fluid is assumed to be infinite in all directions. The model is not intended as the description of an actual macroscopic organism. Rather it serves as a paradigm which is sufficiently simple that the mechanism of swimming can be studied in mathematical detail, at least for small amplitude of stroke.

It is assumed that the entire motion is periodic. This implies that we can take a kinematic point of view and can disregard the equation of motion of the body. In earlier work \cite{5},\cite{6} we showed that the condition that the body exert no total mean force or torque on the fluid leads to a well-defined mean swimming velocity, the mean being defined as a time-average over a period of the motion. Recently we also evaluated the mean flow pattern for a distorting sphere \cite{7}. The amplitude of stroke is assumed to be small in comparison with the size of the body and the theory is based on a perturbation expansion with the ratio of amplitude and size as small parameter. In the model we assume that to first order in this ratio the flow is irrotational. The fluid motion is driven by a pressure wave generated by undulations of the body surface.

Specifically we assume that the surface undulations are axisymmetric and characterized by a dipole and a quadrupole moment. We consider also a model where the surface modulations are characterized by a quadrupole and an octupole moment.  For the dipole-quadrupole model we saw that the mean second order flow pattern is not irrotational, but has an interesting vortex structure, consisting of three vortex rings, two symmetric fore and aft of the body, and a central one of opposite vorticity \cite{7}. Below we find the same property for the quadrupole-octupole model. The calculations suggest that the vortex flow is a general feature of the mean flow pattern in periodic swimming at small amplitude.

It turns out that in the model the mean swimming velocity and mean flow pattern are independent of the kinematic viscosity of the fluid. This provides an interesting connection to the theory of low Reynolds number swimming of micro-organisms \cite{8},\cite{9}. For a sphere of radius $a$ a dimensionless scale number may be defined as $s=a\sqrt{\omega\rho/2\eta}$, where $\omega$ is the frequency, $\rho$ is the mass density of the fluid, and $\eta$ is the shear viscosity. The kinematic viscosity is $\nu=\eta/\rho$. The Stokes limit of low Reynolds number corresponds to $s=0$.

The present model is special, because the complete time-dependent flow to second order in the amplitude parameter can be obtained. The model calculations are performed in the full range of scale number $s$. The inertia-dominated limit corresponds to $s\rightarrow\infty$. For a bird of size $a=10\;cm$ in air the scale number is approximately $s=46$ for $\omega=2\pi\;Hz$. The second harmonic flow depends on the kinematic viscosity of the fluid.

In earlier work we studied the mean swimming velocity and the mean rate of dissipation \cite{10},\cite{11}, as well as the mean flow pattern \cite{12} in the Stokes limit for a more general axisymmetric stroke of small amplitude. Later we calculated these quantities \cite{6},\cite{7} for arbitrary values of the scale number $s$. Ishimoto \cite{13} had studied corrections to the mean swimming velocity and the mean rate of dissipation near the friction-dominated Stokes limit to order $s^2$.

Spelman and Lauga \cite{14} studied the translational velocity of a squirmer in the inertia-dominated limit by the method of matched asymptotic expansion. Wang and Ardekani \cite{15} studied the effect of fluid inertia on the swimming of small organisms via an approximate equation of motion.

Khair and Chisholm \cite{16} calculated the mean swimming velocity of a spherical squirmer to second order in the Reynolds number, and the mean flow to first order, where the Reynolds number is defined by the swimming velocity in the Stokes limit. Chisholm et al. \cite{17} performed numerical calculations for the squirmer across a wide range of Reynolds number. In these articles the squirmer is viewed as an active particle.

\section{\label{II}Dipole-quadrupole swimmer}

We consider a flexible sphere of radius $a$ immersed in a viscous
incompressible fluid of shear viscosity $\eta$ and mass density $\rho$.
The fluid is set in motion by time-dependent distortions of the
sphere. We study in particular a dipole-quadrupole swimmer with axisymmetric periodic distortions leading to translational swimming
in the $z$ direction in a Cartesian system of coordinates.
The prescribed surface displacement is written as \cite{5}
\begin{equation}
\label{2.1}
\vc{\xi}(\theta,t)=\mathrm{Re}[\vc{\xi}_\omega(\theta)e^{-i\omega t}],
\end{equation}
with polar angle $\theta$ and complex amplitude $\vc{\xi}_\omega(\theta)$. The corresponding first order flow velocity and pressure are given by
\begin{equation}
\label{2.2}
\vc{v}^{(1)}(\vc{r},t)=\mathrm{Re}[\vc{v}_\omega(\vc{r})e^{-i\omega t}],\qquad p^{(1)}(\vc{r},t)=\mathrm{Re}[p_\omega(\vc{r})e^{-i\omega t}],
\end{equation}
with amplitude functions which satisfy the linearized Navier-Stokes equations
\begin{equation}
\label{2.3}\eta[\nabla^2\vc{v}_\omega-\alpha^2\vc{v}_\omega]-\nabla p_\omega=0,\qquad\nabla\cdot\vc{v}_\omega=0,
\end{equation}
with the variable
\begin{equation}
\label{2.4}\alpha=(-i\omega\rho/\eta)^{1/2}=(1-i)(\omega\rho/2\eta)^{1/2}.
\end{equation}

We choose in particular a surface displacement $\vc{\xi}(\theta,t)$ for which the first order flow velocity, corresponding to the no-slip boundary condition, is irrotational. This particular choice of stroke has the advantage that also the flow to second order in the amplitude can be found in explicit detail.
For the dipole-quadrupole swimmer the surface distortion with amplitude $\varepsilon$ is given by
\begin{equation}
\label{2.5}\vc{\xi}(\vc{s},t)=\varepsilon a\big[\mu_1\vc{B}_1(\theta)\sin(\omega t)-\mu_2\;\vc{B}_2(\theta)\cos(\omega t)\big],
\end{equation}
where $\vc{s}=a\hat{\vc{r}}$ denotes a point on the undistorted sphere corresponding to polar angle $\theta$, and $\vc{B}_1$ and $\vc{B}_2$ are vector spherical harmonics defined by
 \begin{equation}
\label{2.6}\vc{A}_l=lP_l\vc{e}_r+\frac{\partial P_l}{\partial\theta}\;\vc{e}_\theta,\qquad
\vc{B}_l=-(l+1)P_l\vc{e}_r+\frac{\partial P_l}{\partial\theta}\;\vc{e}_\theta,
\end{equation}
with unit vectors $\vc{e}_r=\hat{\vc{r}},
\;\vc{e}_\theta$, and Legendre polynomials $P_l(\cos\theta)$ in the notation of Edmonds \cite{18}.
The factors $\mu_1,\mu_2$ in Eq. (2.5) are real numbers which can be freely chosen. In the figures below we choose $\mu_1=1,\;\mu_2=1/\sqrt{2}$ such that the efficiency, given by the ratio of swimming speed and rate of dissipation, is optimal \cite{7},\cite{11}. In Fig. 1 we show the shape of the swimmer for amplitude factor $\varepsilon=0.1$ at sixteen equidistant instants of time in a period. The factor $\varepsilon=0.1$ is chosen for graphical clarity. In Fig. 2 we plot $\xi_r(\theta,t)\sin\theta/(\varepsilon a)$ as a function of $\theta$ and $t$ for two periods. In Fig. 3 we plot $\xi_\theta(\theta,t)\sin\theta/(\varepsilon a)$ as a function of $\theta$ and $t$ for two periods. These plots demonstrate the wavelike nature of the surface disturbance. The first order flow corresponding to Eq. (2.5) is \cite{7}
\begin{eqnarray}
\label{2.7}\vc{v}^{(1)}(\vc{r},t)&=&\varepsilon a\omega\bigg[\mu_1\frac{a^3}{r^3}\;\vc{B}_1(\theta)\cos(\omega t)+\mu_2\frac{a^4}{r^4}\;\vc{B}_2(\theta)\sin(\omega t)\bigg],\nonumber\\
p^{(1)}(\vc{r},t)&=&\varepsilon\rho a^2\omega^2\bigg[\mu_1\frac{a^2}{r^2}\;P_1(\cos\theta)\sin(\omega t)-\mu_2\frac{a^3}{\;r^3}\;P_2(\cos\theta)\cos(\omega t)\bigg],
\end{eqnarray}
The integral of $\vc{\xi}(\vc{s},t)$ over the sphere vanishes, so that to first order the center of the sphere is at rest at the origin. The flow velocity satisfies the no-slip boundary condition to first order in the amplitude, implying $\vc{v}^{(1)}(\vc{s},t)=\partial\vc{\xi}(\vc{s},t)/\partial t$. Because the first order flow in Eq. (2.7) is potential, it is independent of fluid viscosity.

The second order velocity field $\vc{v}^{(2)}(\vc{r},t)$ and pressure $p^{(2)}(\vc{r},t)$ satisfy the equations \cite{5}
\begin{equation}
\label{2.8}\rho\bigg(\frac{\partial\vc{v}^{(2)}}{\partial t}+(\vc{v}^{(1)}\cdot\nabla)\vc{v}^{(1)}\bigg)=\eta\nabla^2\vc{v}^{(2)}-\nabla p^{(2)},\qquad\nabla\cdot\vc{v}^{(2)}=0.
\end{equation}
Since the first order flow velocity is irrotational the Reynolds force density $-\rho(\vc{v}^{(1)}\cdot\nabla)\vc{v}^{(1)}$ may be expressed as the gradient of a scalar function which may be identified with a second order pressure deviation $p_B^{(2)}=-\frac{1}{2}\rho\vc{v}^{(1)2}$ of Bernoulli type \cite{5}. As a consequence the Reynolds flow velocity vanishes and the second order flow velocity has only a surface contribution.
It must be found as the solution of the homogenous equations
\begin{equation}
\label{2.9}\rho\frac{\partial\vc{v}^{(2)}}{\partial t}=\eta\nabla^2\vc{v}^{(2)}-\nabla p_S^{(2)},\qquad\nabla\cdot\vc{v}^{(2)}=0,
\end{equation}
where $p^{(2)}_S(\vc{r},t)$ is the remaining second order pressure deviation. The flow velocity $\vc{v}^{(2)}(\vc{r},t)$ must satisfy the second order no-slip boundary condition \cite{5}
\begin{equation}
\label{2.10}\vc{v}^{(2)}(\vc{s},t)=\vc{u}_S(\vc{s},t),\qquad\vc{u}_S(\vc{s},t)=-(\vc{\xi}\cdot\nabla)\vc{v}^{(1)}\big|_{r=a},
\end{equation}
and must tend to $-\overline{U_2}\vc{e}_z$ as $r\rightarrow\infty$, where $\overline{U_2}$ is the mean swimming velocity.

The surface velocity $\vc{u}_S(\vc{s},t)$ can be calculated by use of the expression for $\nabla\vc{v}$ in spherical coordinates given by Happel and Brenner \cite{19}.
As a function of time the surface velocity $\vc{u}_S(\vc{s},t)$ is a sum of zeroth and second order harmonics, and can therefore be expressed as a constant term $\overline{\vc{u}}_S(\vc{s})$ and two terms proportional to $\cos(2\omega t)$ and $\sin(2\omega t)$, respectively.
We calculated the mean surface velocity $\overline{\vc{u}}_S(\vc{s})$ and the corresponding net flow velocity $\vc{v}_{12}^\prime(\vc{r})$ in earlier work \cite{7}. Alternatively we can use the above expressions. The mean surface velocity is given by
\begin{equation}
\label{2.11}\overline{\vc{u}}_S(\vc{s})=\overline{U_2}\bigg[-\vc{A}_1-\frac{4}{5}\vc{B}_1-\frac{2}{35}\vc{A}_3-\frac{1}{7}\vc{B}_3\bigg].
\end{equation}
with mean swimming velocity
\begin{equation}
\label{2.12}\overline{U_2}=3\varepsilon^2\mu_1\mu_2a\omega.
\end{equation}
We recall that $\vc{A}_1=\vc{e}_z$ and $\vc{B}_1=\vc{e}_z-3\vc{e}_r\cos\theta$. It is convenient to use complex notation and write
\begin{equation}
\label{2.13}\vc{v}^{(2)}(\vc{r},t)=-\overline{U_2}\vc{e}_z+\vc{v}_{12}^\prime(\vc{r})+2\;\mathrm{Re}[\vc{v}^{(2)}_-(\vc{r})e^{-2i\omega t}].
\end{equation}

In order to visualize the flow it is convenient to derive it from a Stokes stream function $\psi(r,\theta)$ according to the rule \cite{20}
\begin{equation}
\label{2.14}v_r(r,\theta)=\frac{1}{r^2\sin\theta}\frac{\partial\psi}{\partial\theta},\qquad v_\theta(r,\theta)=\frac{-1}{r\sin\theta}\frac{\partial\psi}{\partial r}.
\end{equation}
Previously we found that the term $\vc{v}_{12}^\prime(\vc{r})$ in Eq. (2.13) can be expressed in the form of Eq. (2.14) with stream function \cite{7}
\begin{equation}
\label{2.15}\psi^{\prime}_{12}(r,\theta)=\frac{1}{16}\;\overline{U_2}\;\frac{a^3}{r^3}\big[3a^2+11r^2+(5a^2-3r^2)\cos2\theta\big]\sin^2\theta.
\end{equation}
By use of the orthonormality relations of the Gegenbauer functions \cite{19} we find that this can be decomposed as
\begin{equation}
\label{2.16}\psi^{\prime}_{12}(r,\theta)=\psi^{\prime p}_{12}(r,\theta)+\psi^{\prime v}_{12}(r,\theta)
\end{equation}
with
\begin{eqnarray}
\label{2.17}\psi^{\prime p}_{12}(r,\theta)&=&\overline{U_2}\;\bigg[\frac{8a^3}{5r}\;\mathcal{I}_2(\cos\theta)
+\frac{a^5}{r^3}\;\mathcal{I}_4(\cos\theta)\bigg],\nonumber\\\psi^{\prime v}_{12}(r,\theta)&=&-\overline{U_2}\;\frac{3a^3}{5r}\;\mathcal{I}_4(\cos\theta).
\end{eqnarray}
The two terms in the first expression are of potential type, satisfying $\nabla^2\nabla^2f=0$, and give rise to irrotational flow. The second expression gives rise to a permanent vortex flow. The steady flow velocity $\vc{v}'_{12}(\vc{r})$ can be expressed as a superposition of modes of the steady state Stokes equations \cite{10},
\begin{equation}
\label{2.18}\vc{v}^{\prime}_{12}(r,\theta)=\overline{U_2}\bigg[\frac{4}{5}\vc{u}_1+\frac{1}{4}\vc{u}_3-\frac{3}{20}\vc{v}^0_3\bigg],
\end{equation}
with
 \begin{equation}
\label{2.19}\vc{u}_l(\vc{r})=-\bigg(\frac{a}{r}\bigg)^{l+2}\vc{B}_l,\qquad
\vc{v}^0_l(\vc{r})=\bigg(\frac{a}{r}\bigg)^l\bigg[\frac{2l+2}{l(2l+1)}\vc{A}_l-\frac{2l-1}{2l+1}\vc{B}_l\bigg].
\end{equation}
It is independent of the shear viscosity and mass density of the fluid. The flow is associated with a pressure disturbance and with vorticity.

The pressure disturbance is given by
\begin{equation}
\label{2.20}p_{12}^\prime(\vc{r})=-\frac{3}{20}\overline{U_2}\;p^0_3(\vc{r}),
\end{equation}
with pressure mode $p^0_l(\vc{r})$ associated to $\vc{v}^0_l(\vc{r})$,
\begin{equation}
\label{2.21}p^0_l(\vc{r})=\eta(4l-2)\frac{a^l}{r^{l+1}}P_l(\cos\theta),
\end{equation}
as defined earlier \cite{7}. The flow $\vc{v}^{\prime}_{12}$ has nonvanishing vorticity
\begin{equation}
\label{2.22}\nabla\times\vc{v}^{\prime}_{12}=-\frac{3}{16}\;\overline{U_2}\;\frac{a^3}{r^4}\big[\sin\theta+5\sin3\theta\big]\vc{e}_\varphi.
\end{equation}
The steady state vorticity is built up in the course of time by the undulating surface via the no-slip boundary condition. The vortex structure is composed of three vortex rings, two of the same vorticity in front and aft of the sphere, and a central one of opposite vorticity.

\section{\label{III}Vortex shedding}

The last term in Eq. (2.11) describes vortex shedding. The second harmonic flow amplitude $\vc{v}^{(2)}_-(\vc{r})$ is complex and can be expressed as a linear superposition of modes \cite{21}
 \begin{eqnarray}
\label{3.1}\vc{v}_l(\vc{r},\beta)&=&\frac{2}{\pi}\;e^{\beta a}[(l+1)k_{l-1}(\beta r)\vc{A}_l(\hat{\vc{r}})+lk_{l+1}(\beta r)\vc{B}_l(\hat{\vc{r}})],\nonumber\\
\vc{u}_l(\vc{r})&=&-\bigg(\frac{a}{r}\bigg)^{l+2}\vc{B}_l(\hat{\vc{r}}),\qquad p_l(\vc{r},\beta)=\eta\beta^2a\bigg(\frac{a}{r}\bigg)^{l+1}P_l(\cos\theta),
\end{eqnarray}
with modified spherical Bessel functions \cite{22} $k_l(z)$, vector spherical harmonics $\{\vc{A}_l,\vc{B}_l\}$, and $\beta=\sqrt{2}\;\alpha$. The superposition coefficients can be found from the expansion of the second harmonic surface velocity
\begin{equation}
\label{3.2}\vc{u}_{S-}(\vc{s})=\frac{1}{T}\int_0^T\exp(2i\omega t)\vc{u}_{S}(\vc{s},t)\;dt,
\end{equation}
where $T=2\pi/\omega$ is the period, in vector spherical harmonics.

By use of the orthonormality relations for the vector spherical harmonics \cite{7} we find that the expansion of the second harmonic surface velocity takes the form
\begin{equation}
\label{3.3}\vc{u}_{S-}(\vc{s})=\varepsilon^2\mu_1\mu_2a\omega\sum^4_{l=1}\big[u_{Al}\vc{A}_l+u_{Bl}\vc{B}_l\big],
\end{equation}
with superposition coefficients
\begin{eqnarray}
\label{3.4}u_{A1}&=&-\frac{3}{2},\qquad u_{A2}=-\frac{9i\sqrt{2}}{140},\qquad u_{A3}=-\frac{9}{35},\qquad u_{A4}=-\frac{3i\sqrt{2}}{56},\nonumber\\
u_{B1}&=&\frac{12}{5},\qquad u_{B2}=\frac{i\sqrt{2}}{35},\qquad u_{B3}=\frac{6}{7},\qquad u_{B4}=\frac{3i\sqrt{2}}{14}.
\end{eqnarray}
From Eq. (3.1) we find that the second harmonic flow velocity has the expansion
\begin{equation}
\label{3.5}\vc{v}^{(2)}_{-}(\vc{r})=\varepsilon^2\mu_1\mu_2a\omega\bigg[\sum^4_{l=1}u_{Al}\frac{\pi e^{-\beta a}}{2(l+1)k_{l-1}(\beta a)}\;\vc{v}_l(\vc{r},\beta)
-\sum^4_{l=1}(u_{Bl}+u'_{Bl})\;\vc{u}_l(\vc{r})\bigg],
\end{equation}
with additional coefficients $\{u'_{Bl}\}$ given by
\begin{eqnarray}
\label{3.6}u'_{B1}&=&\frac{3}{4}\frac{k_2(w)}{k_0(w)},\qquad
u'_{B2}=\frac{3i\sqrt{2}}{70}\frac{k_3(w)}{k_1(w)},\qquad w=\beta a,\nonumber\\
u'_{B3}&=&\frac{27}{140}\frac{k_4(w)}{k_2(w)},\qquad
u'_{B4}=\frac{3i\sqrt{2}}{70}\frac{k_5(w)}{k_3(w)}.
\end{eqnarray}
The coefficients $\{u'_{Bl}\}$ are ratios of polynomials in the variable $w=\beta a$.
By substitution into Eq. (2.10) we find a damped running wave of vortex rings. From the coefficient $u_{A1}$ we determine the time-dependent swimming velocity as \cite{21}
\begin{equation}
\label{3.7}U_2(t)=\overline{U_2}\;[1-\cos(2\omega t)].
\end{equation}

The stream function $\psi^{(2)}_-(r,\theta)$ corresponding to Eq. (3.5) can be expressed as the expansion
\begin{equation}
\label{3.8}\psi^{(2)}_-(r,\theta)=\overline{U_2}\sum^4_{l=1}g_l(r,\beta)\mathcal{I}_{l+1}(\cos\theta).
\end{equation}
The stream function can be written as the sum of an irrotational and a vortex contribution
\begin{equation}
\label{3.9}\psi^{(2)}_-(r,\theta)=\psi^{(2)p}_{-}(r,\theta)+\psi^{(2)v}_{-}(r,\theta),
\end{equation}
where each term has an expansion of the form Eq. (3.8),
\begin{eqnarray}
\label{3.10}\psi^{(2)p}_{-}(r,\theta)&=&\overline{U_2}\;a^2\sum^4_{l=1}c_{pl}\;\bigg(\frac{a}{r}\bigg)^l\;\mathcal{I}_{l+1}(\cos\theta),\nonumber\\
\psi^{(2)v}_{-}(r,\theta)&=&\overline{U_2}\;\beta^{-2}\sum^4_{l=1}c_{vl}\;\frac{\beta rk_l(\beta r)}{k_{l-1}(\beta a)}\;\mathcal{I}_{l+1}(\cos\theta).
\end{eqnarray}
The coefficients $\{c_{pl}\}$ are given by
\begin{eqnarray}
\label{3.11}c_{p1}&=&-\frac{8}{5}-\frac{1}{2}\;\frac{k_2(w)}{k_0(w)},\qquad c_{p2}=-\frac{i\sqrt{2}}{35}-\frac{3i}{35\sqrt{2}}\;\frac{k_3(w)}{k_1(w)},\nonumber\\
c_{p3}&=&-\frac{8}{7}-\frac{9}{35}\;\frac{k_4(w)}{k_2(w)},\qquad
c_{p4}=-\frac{5i}{7\sqrt{2}}-\frac{i}{7\sqrt{2}}\;\frac{k_5(w)}{k_3(w)},
\end{eqnarray}
and the coefficients $\{c_{vl}\}$ are given by
\begin{equation}
\label{3.12}c_{v1}=\frac{3}{2},\qquad c_{v2}=\frac{3i}{7\sqrt{2}},\qquad
c_{v3}=\frac{9}{5},\qquad c_{v4}=\frac{9i}{7\sqrt{2}}.
\end{equation}

The first line in Eq. (3.10) contributes four potentials oscillating at frequency $2\omega$ to the second order stream function $\psi^{(2)}(\vc{r},t)$. The second line contributes four vortex waves. These terms describe vortex shedding. Vorticity is generated at the undulating surface and diffuses into the fluid. The last term in Eq. (2.10) depends strongly on the kinematic viscosity of the fluid. In the limit of small viscosity the second term in Eq. (3.9) differs from zero only in a thin boundary layer, whereas the first term gives rise to irrotational waves of long range.

It is of interest to show various contributions to the flow pattern as a function of time. We choose again $\varepsilon=0.1,\mu_1=1,\mu_2=1/\sqrt{2}$. In Fig. 4 we show the total second order flow pattern in units such that $a=1,\;\omega=1,\;\eta/\rho=1$ at sixteen equidistant times in the first half of a period $T=2\pi/\omega$. We note that the surface displacement in Eq. (2.5) is not invariant under reflection in the $xy$ plane. In Fig. 5 we show the vortex contribution to these flows.
In Fig. 6 we show the value of the vorticity $(\nabla\times\vc{v}^{(2)}(\vc{r},t))\cdot\vc{e}_\varphi/(\varepsilon^2\omega)$ at the surface $r=a$ as a function of $\theta$ during a period.

\section{\label{IV}Force on distant fluid}

In this section we consider the first order flow situation of the dipole-quadrupole model in some more detail. The pressure field in Eq. (2.7) due to the dipole moment is of long range, falling off with distance as $1/r^2$. This implies that the dipolar pressure contributes a flow of momentum which is the same for any sphere of radius $b>a$ centered at the origin. The total flow of momentum across such a spherical surface oscillates in time and is in the $z$ direction. This flow of momentum is caused by the dipolar distortion of the sphere and the corresponding instantaneous displacement of fluid due to incompressibility. There is also momentum flow due to the viscous stress tensor, but its integral over a spherical surface vanishes.

It is instructive to compare the situation with the flow field of a rigid sphere oscillating about the origin due to an applied force $\vc{E}(t)=\mathrm{Re}[\vc{E}_\omega\exp(-i\omega t)]$ acting in the $z$ direction. The force generates a first order velocity given by
\begin{equation}
\label{4.1}\vc{U}^{(1)}(t)=\mathrm{Re}[\vc{U}^{(1)}_\omega\;e^{-i\omega t}],\qquad \vc{U}^{(1)}_\omega=\mathcal{Y}_t(\omega)\vc{E}_\omega,
\end{equation}
where $\mathcal{Y}_t(\omega)$ is the translational admittance of a sphere with no-slip boundary condition \cite{23},
\begin{equation}
\label{4.2}\mathcal{Y}_t(\omega)=\frac{1}{-i\omega(m_0+\frac{1}{2}m_f)+6\pi\eta a(1+\alpha a)},
\end{equation}
where $m_0$ is the mass of the sphere and $m_f=4\pi\rho a^3/3$ is the mass of fluid displaced by the sphere. The complex velocity field is \cite{21}
\begin{equation}
\label{4.3}\vc{v}_{U\omega}(\vc{r})=U^{(1)}_\omega\bigg[\frac{1}{2\alpha^2a^2}(3+3\alpha a+\alpha^2a^2)\vc{u}_1(\vc{r})+\frac{1}{2}\alpha a\;\vc{v}_1(\vc{r},\alpha)\bigg].
\end{equation}
It is checked by use of Eq. (3.1) that
\begin{equation}
\label{4.4}\vc{v}_{U\omega}(\vc{r})\bigg|_{r=a}=U^{(1)}_\omega\vc{e}_z.
\end{equation}
The corresponding complex pressure disturbance is
\begin{equation}
\label{4.5}p_{U\omega}(\vc{r})=\eta U^{(1)}_\omega(3+3\alpha a+\alpha^2a^2)\frac{a}{2r^2}\cos\theta,
\end{equation}
showing the same behavior as in Eq. (2.7), but with a different coefficient.

In assuming the validity of Eqs. (2.5) and (2.7) for surface displacement and flow we consider a situation where an oscillating force is exerted on the fluid at large distance, which is similar to that for an oscillating force acting on a rigid sphere. The force exerted on the body by the fluid can be calculated, but is not needed in the present study. In the next two sections we consider the quadrupole-octupole model, for which the pressure decays faster at infinity, so that there is no momentum flow at large distance.

\section{\label{V}Quadrupole-octupole model}

For the quadrupole-octupole swimmer the surface distortion with amplitude $\varepsilon$ is given by
\begin{equation}
\label{5.1}\vc{\xi}(\vc{s},t)=\varepsilon a\big[\mu_2\vc{B}_2(\theta)\sin(\omega t)-\mu_3\vc{B}_3(\theta)\cos(\omega t)\big],
\end{equation}
with $\vc{B}_2$ and $\vc{B}_3$ given by Eq. (2.6).
In the figures below we choose the prefactors $\mu_2=1,\mu_3=\sqrt{3/5}$ corresponding to optimal mean swimming velocity \cite{11}. In Fig. 7 we show the shape of the swimmer for amplitude factor $\varepsilon=0.1$ at sixteen equidistant instants of time in a period. In Fig. 8 we plot $\xi_r(\theta,t)\sin\theta$ as a function of $\theta$ and $t$ for two periods. In Fig. 9 we plot $\xi_\theta(\theta,t)\sin\theta$ as a function of $\theta$ and $t$ for two periods. The first order flow corresponding to Eq. (5.1) is
\begin{eqnarray}
\label{5.2}\vc{v}^{(1)}(\vc{r},t)&=&\varepsilon a\omega\bigg[\mu_2\frac{a^4}{r^4}\;\vc{B}_2(\theta)\cos(\omega t)+\mu_3\frac{a^5}{r^5}\;\vc{B}_3(\theta)\sin(\omega t)\bigg],\nonumber\\
p^{(1)}(\vc{r},t)&=&\varepsilon\rho a^2\omega^2\bigg[\mu_2\frac{a^3}{r^3}\;P_2(\cos\theta)\sin(\omega t)-\mu_3\frac{a^4}{r^4}\;P_3(\cos\theta)\cos(\omega t)\bigg].
\end{eqnarray}

The mean surface velocity $\overline{\vc{u}}_S(\vc{s})$ is found to be given by
\begin{equation}
\label{5.3}\overline{\vc{u}}_S(\vc{s})=-\overline{U_2}\big[\vc{A}_1+\frac{5}{63}\vc{A}_3+\frac{1}{77}\vc{A}_5+\frac{5}{7}\vc{B}_1+\frac{1}{7}\vc{B}_3+\frac{5}{99}\vc{B}_5\big],
\end{equation}
with mean swimming velocity
\begin{equation}
\label{5.4}\overline{U_2}=6\varepsilon^2\mu_2\mu_3a\omega.
\end{equation}
The corresponding net flow velocity $\vc{v}_{23}^\prime(\vc{r})$ can be expressed as a superposition of modes of the steady state Stokes equations given by Eq. (2.18).
 From Eq. (5.3) we find
 \begin{equation}
\label{5.5}\vc{v}_{23}^\prime(\vc{r})=\overline{U_2}\bigg[\frac{5}{7}\;\vc{u}_1+\frac{7}{24}\;\vc{u}_3+\frac{25}{252}\;\vc{u}_5
-\frac{5}{24}\;\vc{v}^0_3-\frac{5}{84}\;\vc{v}^0_5\bigg].
\end{equation}
The flow fields $\vc{u}_1$ and $\vc{v}^0_3$ both decay as $1/r^3$ at large distance. The first three terms in Eq. (5.5) represent irrotational flows with vanishing pressure disturbance. The last two terms are associated with a pressure disturbance and vorticity. The pressure disturbance is given by
 \begin{equation}
\label{5.6}p_{23}^\prime(\vc{r})=\overline{U_2}\bigg[-\frac{5}{24}p^0_3-\frac{5}{84}p^0_5\bigg].
\end{equation}
The vorticity is given by
\begin{equation}
\label{5.7}\nabla\times\vc{v}^{\prime}_{23}=-\frac{5a^3}{2688r^6}\;\overline{U_2}\;\big[405a^2+840r^2+28(27a^2+50r^2)\cos2\theta+567a^2\cos4\theta\big]\sin\theta\;\vc{e}_\varphi.
\end{equation}
Again the vortex structure is composed of three vortex rings, two of the same vorticity in front and aft of the sphere, and a central one of opposite vorticity.

We find that the stream function can be decomposed as in Eq. (2.16) with
\begin{eqnarray}
\label{5.8}\psi^{\prime p}_{23}(r,\theta)&=&\overline{U_2}\;\bigg[\frac{10a^3}{7r}\;\mathcal{I}_2(\cos\theta)
+\frac{7a^5}{6r^3}\;\mathcal{I}_4(\cos\theta)+\frac{25a^7}{42r^5}\;\mathcal{I}_6(\cos\theta)\bigg],\nonumber\\
\psi^{\prime v}_{23}(r,\theta)&=&\overline{U_2}\;\bigg[-\frac{5a^3}{6r}\;\mathcal{I}_4(\cos\theta)-\frac{15a^5}{42r^3}\;\mathcal{I}_6(\cos\theta)\bigg].
\end{eqnarray}
In Fig. 10 we show the streamlines of the flow.

\section{\label{VI}Second harmonic}

By use of the orthonormality relations for the vector spherical harmonics \cite{7} we find that the expansion of the second harmonic surface velocity of the quadrupole-octupole model takes the form
\begin{equation}
\label{6.1}\vc{u}_{S-}(\vc{s})=\varepsilon^2\mu_2\mu_3a\omega\sum^6_{l=1}\big[u_{Al}\vc{A}_l+u_{Bl}\vc{B}_l\big],
\end{equation}
with superposition coefficients
\begin{eqnarray}
\label{6.2}\{u_{Al}\}&=&\bigg(-3,\;-\frac{i}{14}\sqrt{\frac{5}{3}},\;-\frac{5}{7},\;-\frac{i}{44}\sqrt{15},\;-\frac{15}{77},\;-\frac{5i}{286}\sqrt{15}\bigg),\nonumber\\
\{u_{Bl}\}&=&\bigg(\frac{30}{7},\;0,\;\frac{12}{7},\;\frac{8i}{385}\sqrt{15},\;\frac{10}{11},\;\frac{40i}{143}\sqrt{\frac{5}{3}}\bigg).
\end{eqnarray}
From Eq. (3.1) we find that the second harmonic flow velocity has the expansion
\begin{equation}
\label{6.3}\vc{v}^{(2)}_{-}(\vc{r})=\varepsilon^2\mu_2\mu_3a\omega\bigg[\sum^6_{l=1}u_{Al}\frac{\pi e^{-\beta a}}{2(l+1)k_{l-1}(\beta a)}\;\vc{v}_l(\vc{r},\beta)
-\sum^6_{l=1}(u_{Bl}+u'_{Bl})\;\vc{u}_l(\vc{r})\bigg],
\end{equation}
with additional coefficients $\{u'_{Bl}\}$ given by
\begin{eqnarray}
\label{6.4}u'_{B1}&=&\frac{3}{2}\;\frac{k_2(w)}{k_0(w)},\qquad
u'_{B2}=\frac{i}{21}\sqrt{\frac{5}{3}}\;\frac{k_3(w)}{k_1(w)},\qquad u'_{B3}=\frac{15}{28}\;\frac{k_4(w)}{k_2(w)},\nonumber\\
u'_{B4}&=&\frac{i}{55}\sqrt{15}\;\frac{k_5(w)}{k_3(w)},\qquad u'_{B5}=\frac{25}{154}\;\frac{k_6(w)}{k_4(w)},\qquad u'_{B6}=\frac{15i}{1001}\sqrt{15}\;\frac{k_7(w)}{k_5(w)}.
\end{eqnarray}
By substitution into Eq. (2.10) we find a damped running wave of vortex rings. From the coefficient $u_{A1}$ we determine the time-dependent swimming velocity as \cite{21}
\begin{equation}
\label{6.5}U_2(t)=\overline{U_2}\;[1-\frac{1}{2}\cos(2\omega t)].
\end{equation}

The stream function can be written as the sum of an irrotational and a vortex contribution as in Eq. (3.9),
where each term has an expansion of the form
\begin{eqnarray}
\label{6.6}\psi^{(2)p}_{-}(r,\theta)&=&\overline{U_2}\;a^2\sum^6_{l=1}c_{pl}\;\bigg(\frac{a}{r}\bigg)^l\;\mathcal{I}_{l+1}(\cos\theta),\nonumber\\
\psi^{(2)v}_{-}(r,\theta)&=&\overline{U_2}\;\beta^{-2}\sum^6_{l=1}c_{vl}\;\frac{\beta rk_l(\beta r)}{k_{l-1}(\beta a)}\;\mathcal{I}_{l+1}(\cos\theta).
\end{eqnarray}
The coefficients $\{c_{pl}\}$ are given by
\begin{eqnarray}
\label{6.7}c_{p1}&=&-\frac{10}{7}-\frac{1}{2}\;\frac{k_2(w)}{k_0(w)},\qquad
c_{p2}=-\frac{i}{42}\sqrt{\frac{5}{3}}\;\frac{k_3(w)}{k_1(w)},\nonumber\\
c_{p3}&=&-\frac{8}{7}-\frac{5}{14}\;\frac{k_4(w)}{k_2(w)},\qquad
c_{p4}=-\frac{4i}{77}\sqrt{\frac{5}{3}}-\frac{i}{22}\sqrt{\frac{5}{3}}\;\frac{k_5(w)}{k_3(w)},\nonumber\\
c_{p5}&=&-\frac{10}{11}-\frac{25}{154}\;\frac{k_6(w)}{k_4(w)},\qquad
c_{p6}=-\frac{140i}{429}\sqrt{\frac{5}{3}}-\frac{5i\sqrt{15}}{286}\;\frac{k_7(w)}{k_5(w)},
\end{eqnarray}
and the coefficients $\{c_{vl}\}$ are given by
\begin{eqnarray}
\label{6.8}c_{v1}&=&\frac{3}{2},\qquad c_{v2}=\frac{5i}{42}\sqrt{\frac{5}{3}},\qquad
c_{v3}=\frac{5}{2},\nonumber\\ c_{v4}&=&\frac{3i\sqrt{15}}{22},\qquad c_{v5}=\frac{25}{14},\qquad c_{v6}=\frac{5i\sqrt{15}}{22}.
\end{eqnarray}

The behavior is similar to that in Eq. (3.10). It is of interest to show various contributions to the flow pattern as a function of time.  We choose again $\varepsilon=0.1,\mu_2=1,\mu_3=\sqrt{3/5}$. In Fig. 11 we show the total second order flow pattern in units such that $a=1,\;\omega=1,\;\eta/\rho=1$ at sixteen equidistant times in the first half of a period $T=2\pi/\omega$. Qualitatively the plots are similar to those in Fig. 4.

\section{\label{VII}Flying}
The same models are also useful for the discussion of bird flight. We assume again infinite space, but with a uniform gravitational field in the vertical direction $\vc{e}_g$, causing a force $\vc{G}=G\vc{e}_g$ on the body, independent of position and distortion. The Stokes velocity of the undistorted sphere is $\vc{U}_G=\vc{G}/(6\pi\eta a)$. The distortions of the sphere define a body axis. To the order of our calculation it suffices to point this in a direction such that the velocity $\vc{U}_G$ is canceled and the body flies in a horizontal direction $\vc{e}_h$ perpendicular to $\vc{e}_g$. This defines the angle of attack \cite{24},\cite{25}
\begin{equation}
\label{7.1}\gamma=\arcsin\frac{U_G}{\overline{U_2}}.
\end{equation}
It is assumed that $U_G<\overline{U_2}$. The net body velocity is
\begin{eqnarray}
\label{7.2}\vc{U}_b&=&\vc{U}_G+\overline{U_2}\big[\cos\gamma\;\vc{e}_h-\sin\gamma\;\vc{e}_g\big]\nonumber\\
&=&\overline{U_2}\cos\gamma\;\vc{e}_h,
\end{eqnarray}
in the horizontal direction. In nature the angle of attack is about $6^\circ$.

We recall that the kinematic viscosity of air is about fifteen times that of water. The effect of air compressibility can be taken into account without difficulty.

\section{\label{VIII}Discussion}

The two models we studied above exhibit two conspicuous features, the steady state vortex structure and the second harmonic vortex shedding. The models have the advantage that they take full account of fluid dynamics, as described by the time-dependent Navier-Stokes equations, and of fluid-body interactions, as incorporated in the  no-slip boundary condition. The surface modulations of the body are assumed to be given, and in this sense the description is kinematic.

The motions of fluid and body are assumed to be periodic and transient effects are not considered. It would be of interest to study transient effects in computer simulation. This would allow one to see how the steady state vortex structure is built up. Presumably the quadupole-octupole model is easier to simulate than the dipole-quadrupole model, since in the first the flow patterns are of shorter range than in the latter.

The analysis is based on perturbation theory. This implies that the flow is assumed to be laminar. At larger amplitude there is a transition to turbulence \cite{25}. Nonetheless the models provide intriguing insight in the phenomenon of swimming and flying. The assumption of irrotational first order flow simplifies the analysis considerably and allows straightforward analytic work. It would be of interest to extend our earlier study of the mean flow \cite{7} to an analysis of time-dependent second order flow for general first order axisymmetric stroke.

\newpage

\newpage

\section*{Figure captions}

\subsection*{Fig. 1}
Shape of the dipole-quadrupole swimmer for amplitude factor $\varepsilon=0.1$ at sixteen equidistant instants of time in a period $T=2\pi/\omega$. The figure is to be read from left to right and from top to bottom. The body swims in the vertical direction, corresponding to the polar axis.

\subsection*{Fig. 2}
Plot of $(\xi_r(\theta,t)\sin\theta)/(\varepsilon a)$ (axis-label $\xi_r$) for the dipole-quadrupole model during two periods of the motion, $0<\omega t<4\pi$, for $0<\theta<\pi$.

\subsection*{Fig. 3}
Plot of $(\xi_\theta(\theta,t)\sin\theta)/(\varepsilon a)$ (axis-label $\xi_\theta$) for the dipole-quadrupole model during two periods of the motion, $0<\omega t<4\pi$, for $0<\theta<\pi$.

\subsection*{Fig. 4}
Second order flow patterns of the dipole-quadrupole swimmer at sixteen equidistant instants of time in the first half of a period $T=2\pi/\omega$ for $a=1,\;\omega=1,\;\eta/\rho=1$. The plots are drawn in the $xz$ plane.

\subsection*{Fig. 5}
Vortex contribution to the flows in Fig. 4. The plots are drawn in the $xz$ plane.

\subsection*{Fig. 6}
Vorticity $c_\varphi=(\nabla\times\vc{v}^{(2)}(\vc{r},t))\cdot\vc{e}_\varphi/(\varepsilon^2\omega)$ at the surface $r=a$ as a function of $\theta$ during a period, $0<\omega t<2\pi$, for $0<\theta<\pi$, for the dipole-quadrupole model with $a=1,\;\omega=1,\;\eta/\rho=1$.

\subsection*{Fig. 7}
Shape of the quadrupole-octupole swimmer for amplitude factor $\varepsilon=0.1$ at sixteen equidistant instants of time in a period $T=2\pi/\omega$. The figure is to be read from left to right and from top to bottom. The body swims in the vertical direction, corresponding to the polar axis.

\subsection*{Fig. 8}
Plot of $(\xi_r(\theta,t)\sin\theta)/(\varepsilon a)$ for the quadrupole-octupole model during two periods of the motion, $0<\omega t<4\pi$, for $0<\theta<\pi$.

\subsection*{Fig. 9}
Plot of $(\xi_\theta(\theta,t)\sin\theta)/(\varepsilon a)$ for the quadrupole-octupole model during two periods of the motion, $0<\omega t<4\pi$, for $0<\theta<\pi$.

\subsection*{Fig. 10}
Streamlines of the steady state flow $\vc{v}'_{23}(\vc{r})$ for the quadrupole-octupole swimmer.

\subsection*{Fig. 11}
Second order flow patterns of the quadrupole-octupole swimmer at sixteen equidistant instants of time in the first half of a period $T=2\pi/\omega$ for $a=1,\;\omega=1,\;\eta/\rho=1$. The plots are drawn in the $xz$ plane.

\newpage
\setlength{\unitlength}{1cm}
\begin{figure}
 \includegraphics{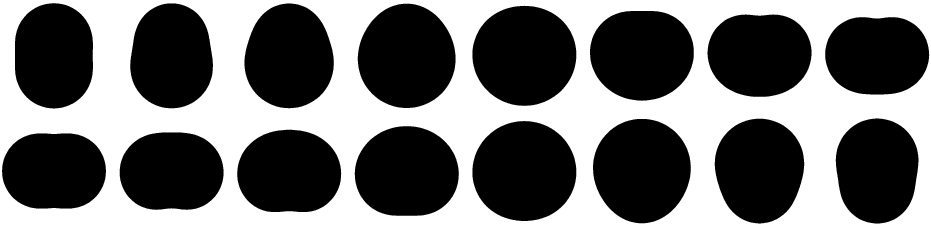}
   \put(-9.1,3.1){}
\put(-1.2,-.2){}
  \caption{}
\end{figure}
\newpage
\clearpage
\newpage
\setlength{\unitlength}{1cm}
\begin{figure}
 \includegraphics{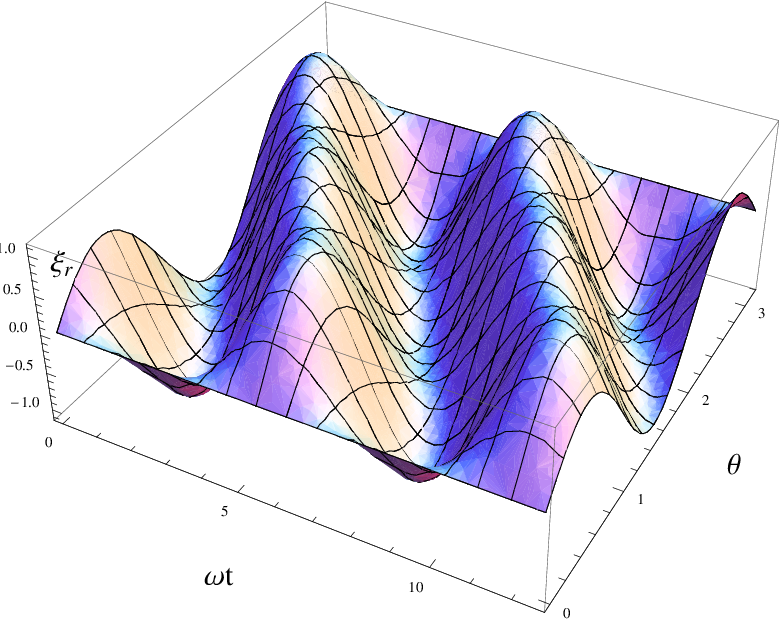}
   \put(-9.1,3.1){}
\put(-1.2,-.2){}
  \caption{}
\end{figure}
\newpage
\clearpage
\newpage
\setlength{\unitlength}{1cm}
\begin{figure}
 \includegraphics{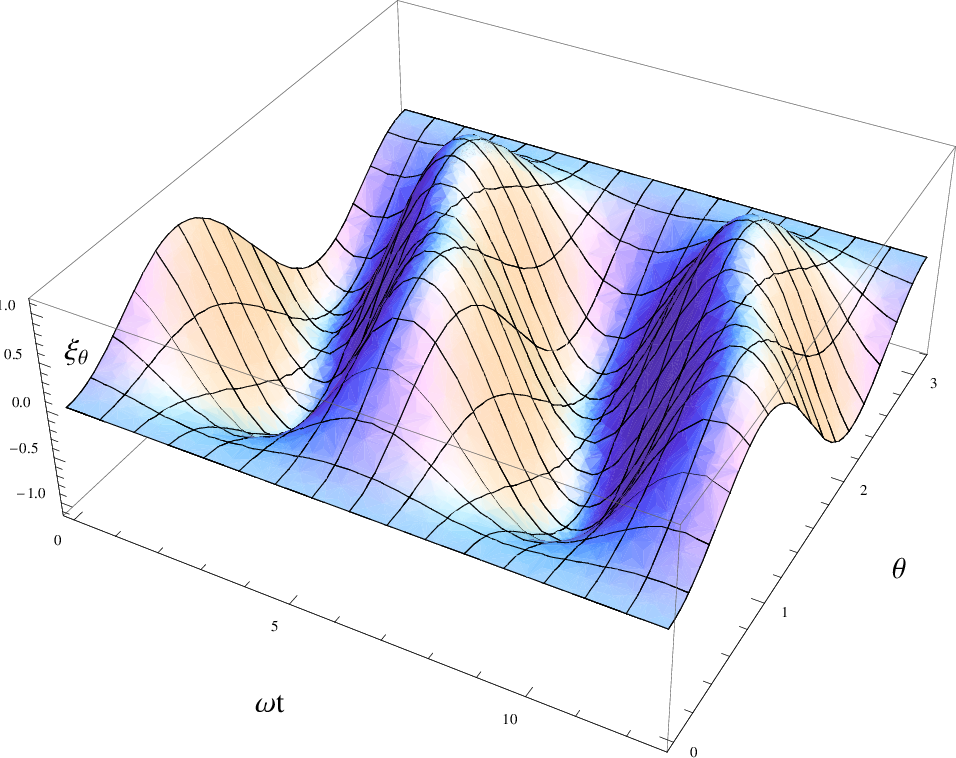}
   \put(-9.1,3.1){}
\put(-1.2,-.2){}
  \caption{}
\end{figure}
\newpage
\clearpage
\newpage
\setlength{\unitlength}{1cm}
\begin{figure}
 \includegraphics{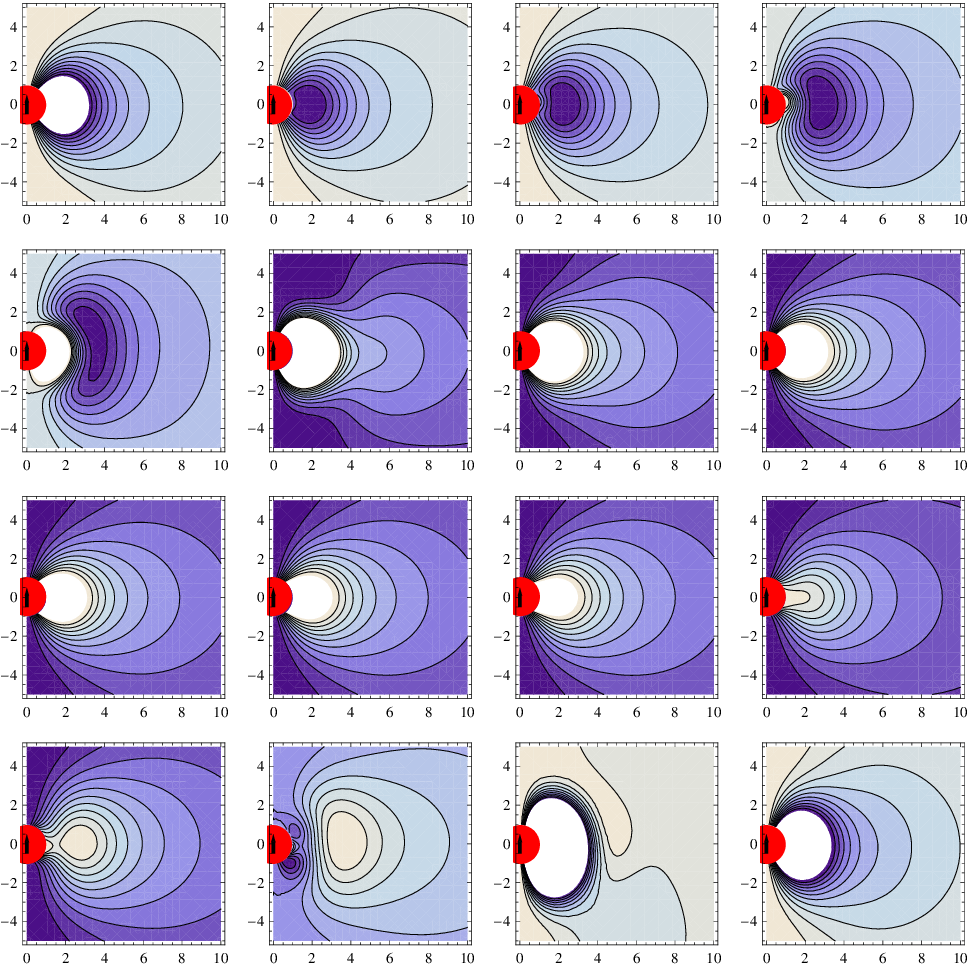}
   \put(-9.1,3.1){}
\put(-1.2,-.2){}
  \caption{}
\end{figure}
\newpage
\clearpage
\newpage
\setlength{\unitlength}{1cm}
\begin{figure}
 \includegraphics{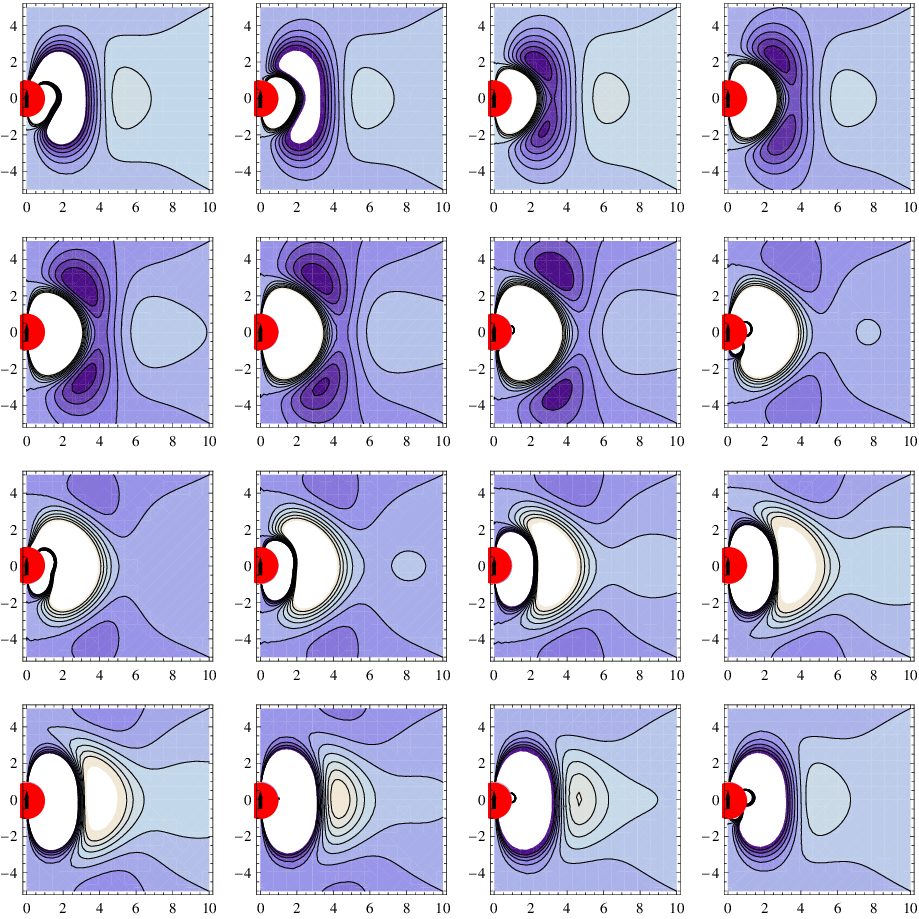}
   \put(-9.1,3.1){}
\put(-1.2,-.2){}
  \caption{}
\end{figure}
\newpage
\clearpage
\newpage
\setlength{\unitlength}{1cm}
\begin{figure}
 \includegraphics{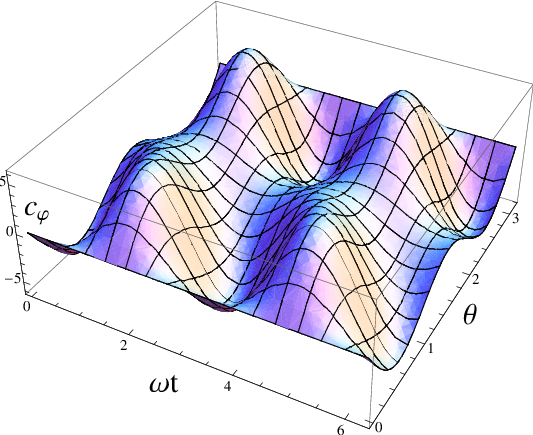}
   \put(-9.1,3.1){}
\put(-1.2,-.2){}
  \caption{}
\end{figure}
\clearpage
\newpage
\newpage
\begin{figure}
 \includegraphics{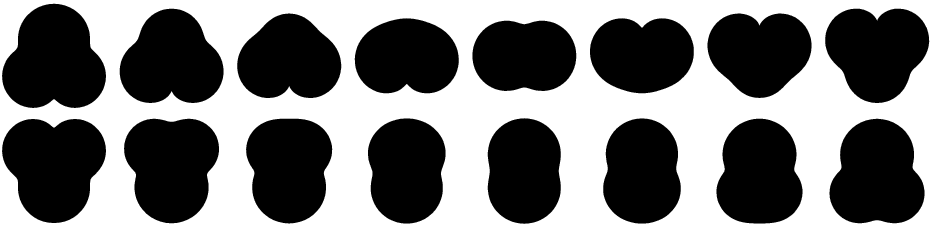}
   \put(-9.1,3.1){}
\put(-1.2,-.2){}
  \caption{}
\end{figure}
\newpage
\clearpage
\newpage
\setlength{\unitlength}{1cm}
\begin{figure}
 \includegraphics{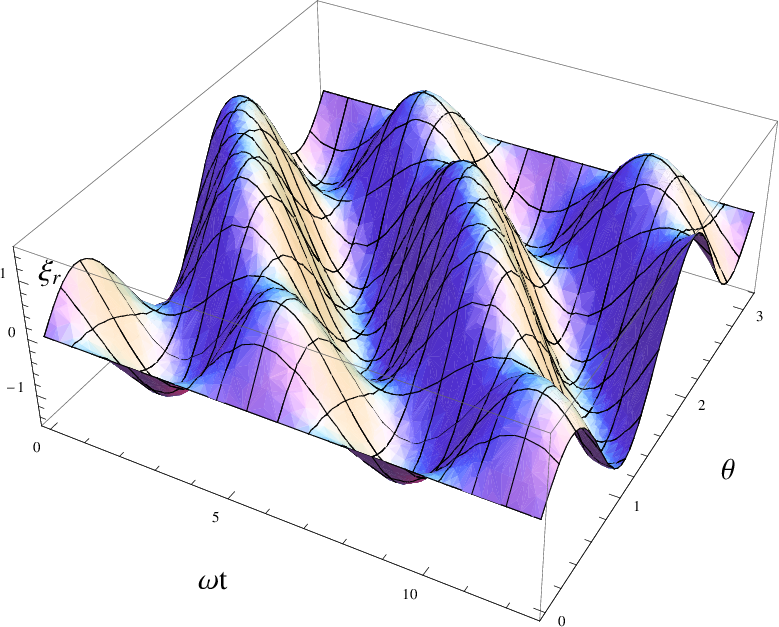}
   \put(-9.1,3.1){}
\put(-1.2,-.2){}
  \caption{}
\end{figure}
\newpage
\clearpage
\newpage
\setlength{\unitlength}{1cm}
\begin{figure}
 \includegraphics{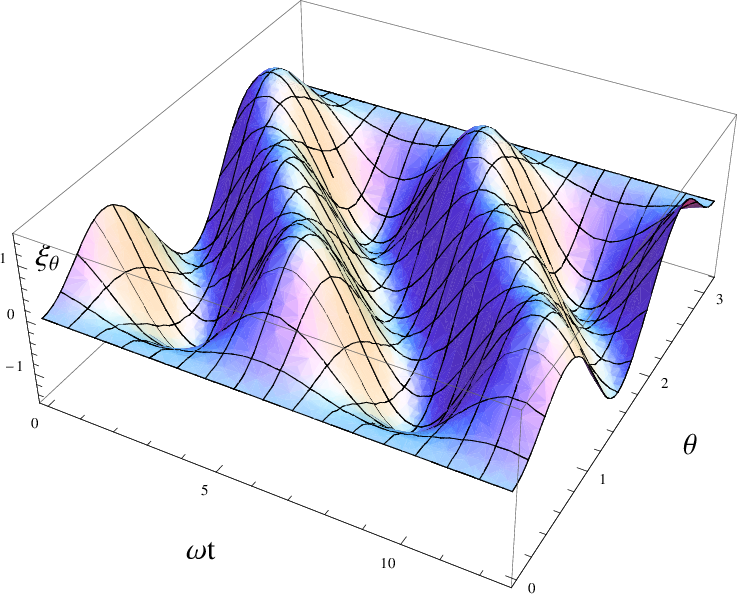}
   \put(-9.1,3.1){}
\put(-1.2,-.2){}
  \caption{}
\end{figure}
\newpage
\clearpage
\newpage
\setlength{\unitlength}{1cm}
\begin{figure}
 \includegraphics{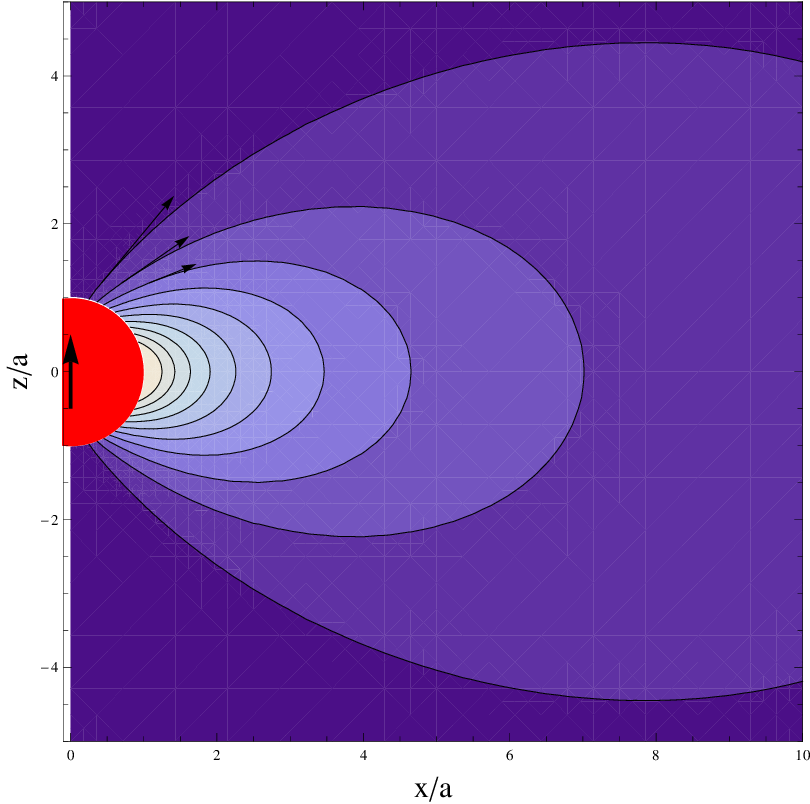}
   \put(-9.1,3.1){}
\put(-1.2,-.2){}
  \caption{}
\end{figure}
\newpage
\clearpage
\newpage
\setlength{\unitlength}{1cm}
\begin{figure}
 \includegraphics{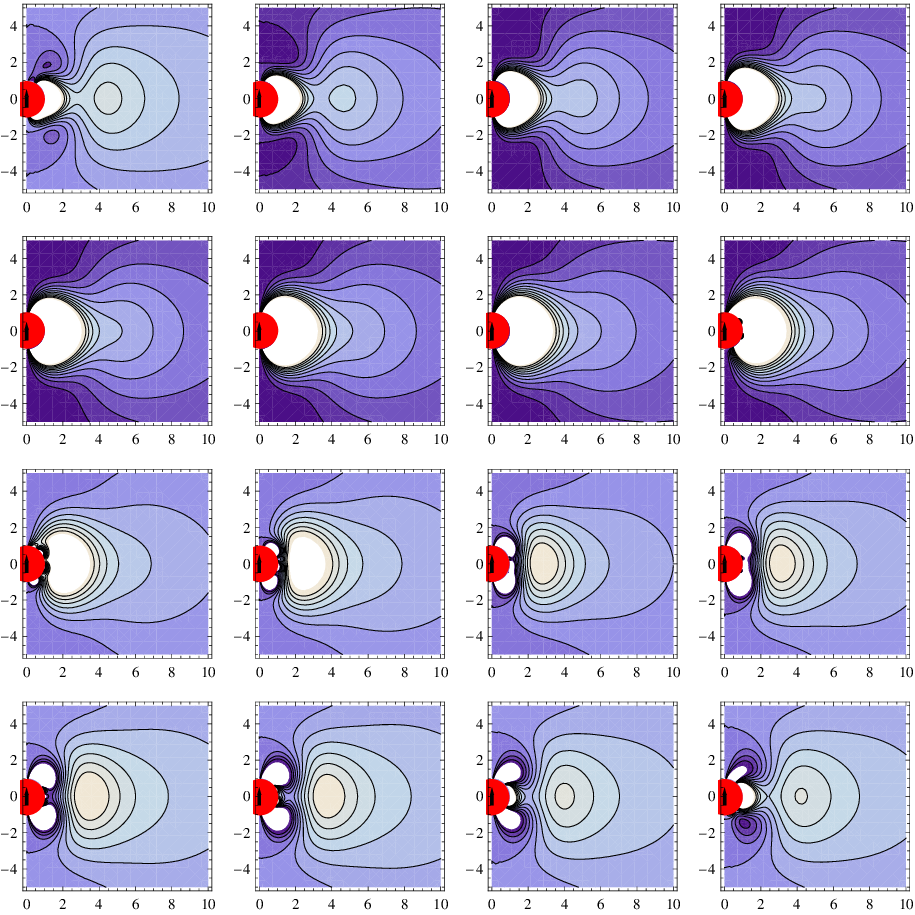}
   \put(-9.1,3.1){}
\put(-1.2,-.2){}
  \caption{}
\end{figure}
\newpage
\end{document}